\documentclass[showpacs,amsmath,amssymb,aps,showkeys,floatfix,prd,a4paper]{revtex4}

\usepackage[dvips]{graphicx}
\usepackage{dcolumn}
\usepackage{bm}
\usepackage{epsfig}
\usepackage{amsfonts}
\usepackage{amssymb,amscd}

\def\lsim{\raise0.3ex\hbox{$<$\kern-0.75em\raise-1.1ex\hbox{$\sim$}}}

\def\gsim{\raise0.3ex\hbox{$>$\kern-0.75em\raise-1.1ex\hbox{$\sim$}}}

\newcommand{\be}{\begin{equation}}

\newcommand{\ee}{\end{equation}}

\def\beq{\begin{equation}}

\def\eeq{\end{equation}}

\def\beqa{\begin{eqnarray}}

\def\eeqa{\end{eqnarray}}

\newcommand{\rd}{\mbox{\boldmath $\Delta$}}

\newcommand{\ba}{\begin{eqnarray}}

\newcommand{\rr}{\mbox{\boldmath $r$}}

\newcommand{\rb}{\mbox{\boldmath $b$}}

\def\gappeq{\mathrel{\rlap {\raise.5ex\hbox{$>$}}

{\lower.5ex\hbox{$\sim$}}}}

\def\lappeq{\mathrel{\rlap{\raise.5ex\hbox{$<$}}

{\lower.5ex\hbox{$\sim$}}}}

\def\Toprel#1\over#2{\mathrel{\mathop{#2}\limits^{#1}}}

\begin{document}

\begin{flushright}
MS-TP-22-38
\end{flushright}

\title{Associated $\phi$ and $J/\Psi$  photoproduction in ultraperipheral $PbPb$ collisions \\ at the Large Hadron Collider and Future Circular Collider}

\author{Celsina N. {\sc Azevedo}}
\email{acelsina@gmail.com}
\affiliation{Institute of Physics and Mathematics, Federal University of Pelotas, \\
  Postal Code 354,  96010-900, Pelotas, RS, Brazil}

\author{Victor P. {\sc Gon\c{c}alves}}
\email{barros@ufpel.edu.br}
\affiliation{Institut f\"ur Theoretische Physik, Westf\"alische Wilhelms-Universit\"at M\"unster,
Wilhelm-Klemm-Straße 9, D-48149 M\"unster, Germany
}
\affiliation{Institute of Modern Physics, Chinese Academy of Sciences,
  Lanzhou 730000, China}
\affiliation{Institute of Physics and Mathematics, Federal University of Pelotas, \\
  Postal Code 354,  96010-900, Pelotas, RS, Brazil}

\author{Bruno D. {\sc Moreira}}
\email{bduartesm@gmail.com}
\affiliation{Departamento de F\'isica, Universidade do Estado de Santa Catarina, 89219-710 Joinville, SC, Brazil.}

\begin{abstract}
In this paper we analyze the associated $\phi$ and $J/\Psi$  photoproduction in ultraperipheral $PbPb$ collisions through the double scattering mechanism for the energies of the Large Hadron Collider (LHC) and Future Circular Collider (FCC). Our results complement a previous analysis for the $\rho\rho$, $J/\Psi J/\Psi$ and $\rho J/\Psi$ production. We present our predictions for the total cross sections and rapidity distributions considering the rapidity ranges covered by the ALICE and LHCb detectors, which indicate that a future experimental analysis of $\phi J/\Psi$ final state is feasible.  These results point out that the study of the double vector meson photoproduction in heavy ion collisions  can be useful to constrain the double scattering mechanism and improve our understanding of the QCD dynamics at high energies.
\end{abstract}



\maketitle

\vspace{1cm}

\section{Introduction}

Over the last two decades, the study of photon -- induced interactions in hadronic colliders became a reality and currently the LHC is also considered a powerful photon -- hadron and photon -- photon collider,  which can be used to  improve our understanding of the standard model as well as to searching for New Physics \cite{upc}.
Since the pioneering works performed in Refs. \cite{klein,gluon,Frankfurt:2001db}, several theoretical and experimental groups have dedicated  its efforts to improve the  description of the exclusive vector meson photoproduction and to measure this process in proton - proton ($pp$), proton -- nucleus ($pA$) and nucleus -- nucleus ($AA$) collisions (For a recent review see, e.g. Ref. \cite{Klein:2020nvu}).  
In general, these studies focused on the production of a single vector meson in the final state. However, the results presented in Ref. \cite{klein} for the vector meson photoproduction in ultraperipheral heavy ion collisions  also has demonstrated that the simultaneous production of two vector mesons is non - negligible. 
Such a conclusion has been confirmed by the detailed studies performed in Refs. \cite{Klusek-Gawenda:2013dka,DSM}, which have compared the predictions associated to the double scattering mechanism (DSM), where two photon -- hadron interactions occur simultaneously, with those derived estimating the double meson production in photon -- photon interactions.  In particular, Ref. \cite{DSM} have demonstrated that the $\rho\rho$ and  $J/\Psi J/\Psi$ production in $PbPb$ collisions is dominated by the double scattering mechanism, while the two - photon mechanism dominates in $pp$ collisions. Moreover, the results presented in that reference have indicated that the analysis of the $\rho J/\Psi$ production at LHC can be useful to constrain the double scattering mechanism.  First experimental results  for the
 exclusive double $J/\Psi$ production in $pp$ collisions  were presented by the  LHCb Collaboration in Ref. \cite{lhcb_dif}, which has   demonstrated that the  analysis of this process is feasible. 
 Data for other combinations of vector mesons in the final state and for heavy ion collisions are currently under analysis and its releasing is expected in the forthcoming years \cite{private}. In particular, the analysis of the associated production of the $\phi$ and  $J/\Psi$ mesons   is being considered, but a theoretical prediction for the corresponding cross section is still not available in the literature.
 Our goal in this paper is to fill this gap and provide, for the first time, the predictions for the $\phi J/\Psi$ production through the double scattering mechanism in heavy ion collisions. As a by - product, we will complement the results presented in Ref. \cite{DSM} by also providing predictions for the double $\phi$ production.  We will consider $PbPb$ collisions for the energies of the LHC ($\sqrt{s} = 5.02$ and 5.5 TeV) and FCC ($\sqrt{s} = 39$ TeV) and    estimate the 
 total cross sections and rapidity distributions for the $\phi J/\Psi$ production assuming the rapidity ranges  covered by the ALICE ($-2.5 \le Y \le 2.5$) and LHCb ($2.0 \le Y \le 4.5$) detectors. As we will demonstrate below, a future experimental analysis of this final state is, in principle, feasible and can be used to improve our understanding of the vector meson production and double scattering mechanism.


\section{Formalism}

\begin{figure}[t]
\includegraphics[scale=0.4]{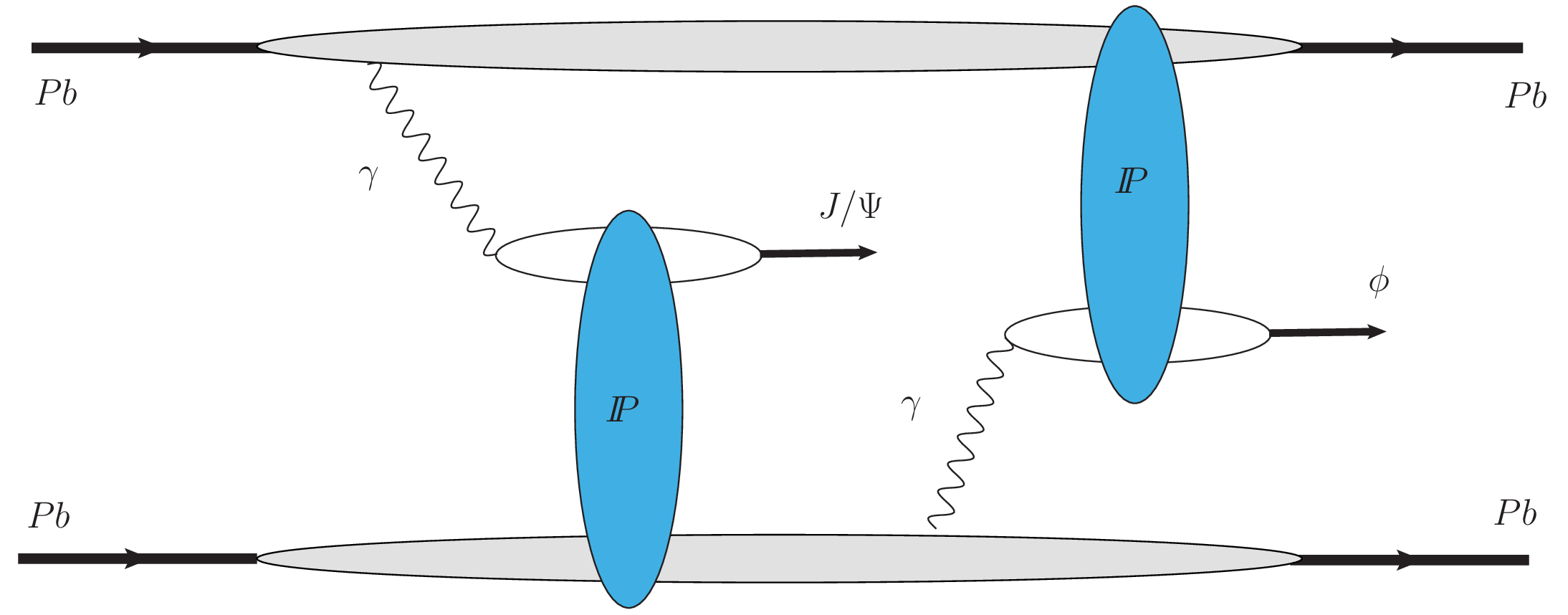}
\caption{One of the four diagrams that contribute for the associated $\phi$ and $J/\Psi$ photoproduction in ultraperipheral heavy ion collisions.}
\label{Fig:diagram}
\end{figure} 

A typical diagram that contributes for the associated $\phi$ and $J/\Psi$ photoproduction in ultraperipheral $PbPb$ collisions through the double scattering mechanism is presented in Fig. \ref{Fig:diagram}. One has that the number of photons emitted by the incident ions is huge ($\approx Z^2$) and that the cross section for the exclusive vector meson photoproduction in a nuclear target is appreciable, which implies that the probability of two simultaneous scatterings becomes non - negligible \cite{klein}. As the formalism to estimate this contribution has been discussed in detail in Ref. \cite{DSM}, in what follows we will only present a brief review of the main formulae and refer the interested reader to that reference for more details. One has that the double differential cross section for the production of a $\phi$ meson at rapidity $y_{\phi}$ and a $J/\Psi$ meson at rapidity $y_{J/\Psi}$ in an ultraperipheral $PbPb$ collision is given by \cite{klein,Klusek-Gawenda:2013dka,DSM}
\begin{eqnarray}
\frac{d^2\sigma \left[Pb Pb  \rightarrow   Pb \,\, \phi \, J/\Psi \,\, Pb \right]}{dy_{\phi} dy_{J/\Psi}} =  \int_{b_{min}}
\frac{d\sigma \,\left[Pb Pb \rightarrow   Pb \otimes \phi \otimes Pb \right]}{d^2b dy_{\phi}}
\times 
\frac{d\sigma \,\left[Pb Pb \rightarrow   Pb \otimes  J/\Psi \otimes Pb \right]}{d^2b dy_{J/\Psi}}
\,\, d^2b \,\,,
\label{Eq:double}
\end{eqnarray}  
where  $b_{min} = 2 R_{Pb}$, { which is equivalent to treat 
the nuclei as hard spheres and that  excludes the overlap between the colliding hadrons, allowing us to take into account only ultraperipheral collisions. {It is important to emphasize that  Eq. (\ref{Eq:double}) can also be estimated assuming  $b_{min} = 0$ and including the survival factor $P_{NH}(b)$ that describes the probability of no additional hadronic interaction between the nuclei, which is usually estimated using the Glauber formalism \cite{klein,Baltz:2009jk}. Such distinct approaches imply different behaviours of the photon spectrum for large photon energies, with the Glauber approach predicting a harder photon spectrum. One has verified that if the Glauber approach is used in our calculations, the impact on our predictions  is smaller than 5\% in the kinematical region of interest (For a more detailed discussion see, e.g., Ref. \cite{Azevedo:2019fyz}). }}  From Eq. (\ref{Eq:double}) one has that the double vector meson production can be  estimated in terms of the cross sections for the single vector meson production. Such quantity is determined by the equivalent photon spectrum associated to the ion, $N_{Pb}(\omega,b)$, and by the  cross section for the $\gamma Pb \rightarrow V Pb$ process, as follows
\begin{eqnarray}
\frac{d\sigma \,\left[Pb Pb \rightarrow   Pb \otimes  V \otimes Pb\right]}{d^2b dy_V} = \omega N_{Pb}(\omega,b)\,\sigma_{\gamma Pb \rightarrow V \otimes Pb}\left(\omega \right)\,\,,
\label{dsigdy}
\end{eqnarray}
where the rapidity $y_V$ of the vector meson in the final state is determined by the photon energy $\omega$ in the collider frame and by mass $M_{V}$ of the vector meson [$y_V\propto \ln \, ( \omega/M_{V})$]. Moreover,  the symbol
$\otimes$ represents the presence of a rapidity gap in the final state. { The photon spectrum can be expressed in terms  in terms of the charge form factor $F(q)$ as follows \cite{upc}
\begin{eqnarray}
 N(\omega,b) = \frac{Z^{2}\alpha}{\pi^2}\frac{1}{b^{2} v^{2}\omega}
\cdot \left[
\int u^{2} J_{1}(u) 
F\left(
 \sqrt{\frac{\left( \frac{b\omega}{\gamma_L}\right)^{2} + u^{2}}{b^{2}}}
 \right )
\frac{1}{\left(\frac{b\omega}{\gamma_L}\right)^{2} + u^{2}} \mbox{d}u
\right]^{2} \,\,,
\label{fluxo}
\end{eqnarray}
where $\alpha$ is the electromagnetic coupling constant, $\gamma_L$ is the Lorentz factor and $v$ is the nucleus velocity. As in previous studies \cite{Klusek-Gawenda:2013dka,DSM}, we will estimate the equivalent photon spectra
for $A = Pb$ assuming the nucleus as a point-like object, i. e. $F(q^2) = 1$. 
 }
For the calculations of the single vector meson photoproduction, we will  take into account that both incident ions can be a source of the photons that interact with the other ion. Consequently, our predictions for the associated $\phi$ and $J/\Psi$ photoproduction in $PbPb$ collisions using Eq. (\ref{Eq:double}) will include the contribution of four terms, one of them represented in Fig. \ref{Fig:diagram}.  
As in our previous study \cite{DSM}, we will estimate $\sigma_{\gamma Pb \rightarrow V \otimes Pb}$ using the color dipole formalism, which predicts that this cross section can be expressed as follows
\begin{eqnarray}
\sigma (\gamma Pb \rightarrow V Pb) =  \int_{-\infty}^0 \frac{d\sigma}{d{t}}\, d{t}  
= \frac{1}{16\pi}  \int_{-\infty}^0 |{\cal{A}}_T^{\gamma Pb \rightarrow V Pb }(x,\Delta)|^2 \, d{t}\,\,,
\label{sctotal_intt}
\end{eqnarray}
with the scattering amplitude being given by
 \begin{eqnarray}
 {\cal A}_{T}^{\gamma Pb \rightarrow V Pb}({x},\Delta)  =  i
\int dz \, d^2\rr \, d^2\rb_A  e^{-i[\rb_A-(1-z)\rr].\rd} 
 \,\, (\Psi^{V*}\Psi)_{T}  \,\,2 {\cal{N}}_{Pb}({x},\rr,\rb_A) \,\,,
\label{sigmatot2}
\end{eqnarray}
where $(\Psi^{V*}\Psi)_{T}$ denotes the overlap of the transverse photon and vector meson wave functions and  $z$ $(1-z)$ is the
longitudinal momentum fractions of the quark (antiquark). As in Ref. \cite{DSM}, the overlap function will be described by the Gaus-LC model, which assumes that the vector meson is predominantly a quark-antiquark state 
and that the spin and polarization structure is the same as in the  photon \cite{dgkp,nnpz,sandapen,KT}, with the parameters presented in Ref. \cite{run2}. Moreover,  $\Delta$ denotes the transverse 
momentum lost by the outgoing ion (${t} = - \Delta^2$) and $\rb_A$ is the transverse distance from the center of the ion to the center of mass of the $q \bar{q}$  dipole.
Finally, ${\cal{N}}_{Pb} (x,\rr,\rb_A)$ denotes the non-forward scattering  amplitude of a dipole of size $\rr$ on the lead ion, which is  directly related to  the QCD dynamics.
Following Ref. \cite{DSM}, we will estimate ${\cal{N}}_{Pb}$  assuming the Glauber-Gribov (GG) formalism~\cite{glauber,gribov,mueller,Armesto:2002ny}, which predicts that
\begin{eqnarray}
{\cal{N}}_A(x,\rr,\rb_A) =  1 - \exp \left[-\frac{1}{2}  \, \sigma_{dp}(x,\rr^2) \,T_A(\rb_A)\right] \,,
\label{enenuc}
\end{eqnarray}
where the nuclear profile function $T_A(\rb_A)$ is described by a Woods-Saxon distribution. The dipole-proton cross section, $\sigma_{dp}$, is expressed in terms of the dipole-proton scattering amplitude as follows
\begin{eqnarray}
\sigma_{dp}(x,\rr^2) = 2 \int \mathrm{d}^2\rb_p \, {\cal{N}}_p(x,\rr,\rb_p) \,,
\end{eqnarray}
with  $\rb_p$ being the impact - parameter for the dipole-proton interaction. In our analysis, the  b-CGC  model for a proton target  \cite{KMW} will be used as input to estimate ${\cal{N}}_A(x,\rr,\rb_A)$.  We will assume the parameters for the b-CGC model obtained in Ref. \cite{amir} considering the high precision combined HERA data, where the authors have demonstrated that this model provides a very good description of the $ep$ HERA data for inclusive and exclusive processes. One has verified that similar predictions are obtained assuming other phenomenological models for ${\cal{N}}_p$.

\begin{figure}[t]
\includegraphics[scale=0.4]{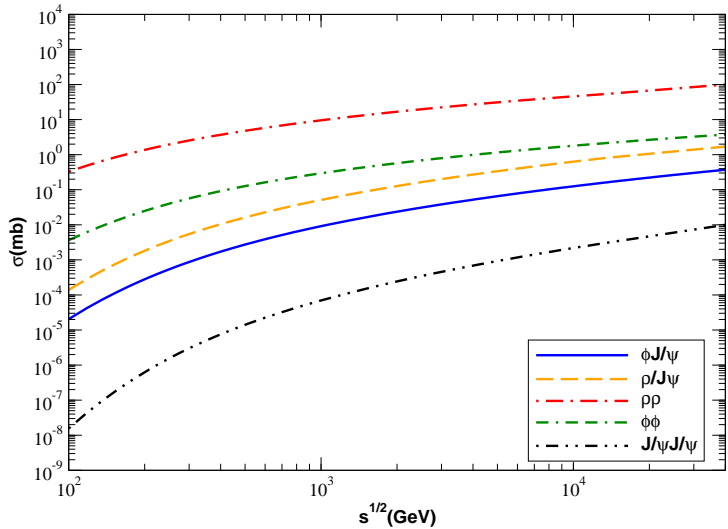}
\caption{Energy dependence of total cross section for the associated $\phi$ and $J/\Psi$ photoproduction in ultraperipheral $PbPb$ collisions (solid line).  The predictions for other combinations of vector mesons in the final state are also presented for comparison.}
\label{Fig:total}
\end{figure}

\section{Results}

Lets start our analysis presenting in Fig. \ref{Fig:total} the predictions for the energy dependence of the total cross section for the associated $\phi$ and $J/\Psi$ photoproduction in ultraperipheral $PbPb$ collisions through the double scattering mechanism. For comparison we also show the predictions for other combinations of vector mesons in the final state derived in Ref. \cite{DSM}.  For completeness of our analysis, we present, for the first time,  the predictions for the double $\phi$ production. One has that the predictions for the $\phi J/\Psi$ production are  approximately one order of magnitude smaller that the results for the $\rho J/\Psi$ case, which is expected due to the larger mass of the $\phi$ meson.  These results indicate that the ideal processes to investigate the double scattering mechanism are  the $\rho \rho$, $\phi \phi$ and $\rho J/\Psi$ final states. The $\rho \rho$ and $\rho J/\Psi$ final states can, in principle, be measured by the ALICE Collaboration, which is able to precisely reconstruct the $\rho$ meson. In contrast, for the LHCb Collaboration, such measurement is still a hard task. Another important aspect is that the exclusive $\rho$ photoproduction is dominated by large dipole sizes \cite{run2}, where nonperturbative contributions are expected to determine the behaviour of the cross section. It implies that the predictions for the $\rho \rho$ and $\rho J/\Psi$ final states are sensitive to the soft regime of the QCD dynamics. In contrast, the photoproduction of $\phi$ and $J/\Psi$ mesons are dominated by smaller dipoles and, consequently, the production of the $\phi \phi$, $\phi J/\Psi$ and $J/\Psi J/\Psi$ final states probes a distinct regime of the QCD dynamics. Therefore, a future measurement of these distinct final states will be very useful to constrain the description of the double scattering mechanism and to improve our understanding of the QCD dynamics. In what follows, we will focus on the $\phi J/\Psi$ final state, which  could be measured by both  ALICE and LHCb Collaborations. Although the cross section for this final state is smaller than for the double $\phi$ production, the presence of the $J/\Psi$ makes its experimental separation more simple to be performed.  
Our predictions for the total cross sections are presented in Table \ref{Tab:secao-de-choque} considering the LHC and FCC energies as well as different rapidity ranges for the mesons in the final state. 
For the LHC energies, one has that the total cross section is reduced by a factor $\approx 2 \, (30)$ when we impose that the mesons are produced in rapidity range covered by the ALICE (LHCb) detector. At the FCC and central rapidities, such suppression becomes larger since the rapidity distribution becomes wider with the increasing of the energy (see below).  Considering the  integrated luminosity expected for future runs of heavy ion collisions at the LHC, ${\cal{L}} \approx 7$ nb$^{-1}$, the associated number of events will be $\ge 10^5 \, (10^4)$ considering the central (forward) rapidity range. Assuming a similar integrated luminosity for the FCC, our predictions for the number of events increase by one order of magnitude.  Such results indicate that a future experimental analysis of this final state is, in principle, feasible.

\begin{table}[t]
\begin{center}

\begin{tabular}{||c|c|c|c||}\hline \hline
                    & \bf{LHC} ($\sqrt{s} = 5.02$ TeV) & \bf{LHC} ($\sqrt{s} = 5.5$ TeV)  & \bf{FCC} ($\sqrt{s} = 39$ TeV)   \\ \hline \hline
Full rapidity range & $65.60 $ & $72.00 $ & $365.00 $ \\ \hline
$-2.5 \le y_{\phi, J/\Psi} \le 2.5$    & $33.60 $ & $35.50$ & $80.30$ \\ \hline
$\;\;2.0 \le y_{\phi, J/\Psi} \le 4.5$ & $2.30$ & $2.70$ & $18.70 $ \\ \hline \hline
\end{tabular} 
\caption{Total cross sections in $\mu$b for the associated $\phi$ and  $J/\Psi$ production in ultraperipheral $PbPb$ collisions at the LHC and FCC energies considering distinct rapidity ranges for the mesons in the final state.}
\label{Tab:secao-de-choque}
\end{center}
\end{table}

\begin{figure}[t]
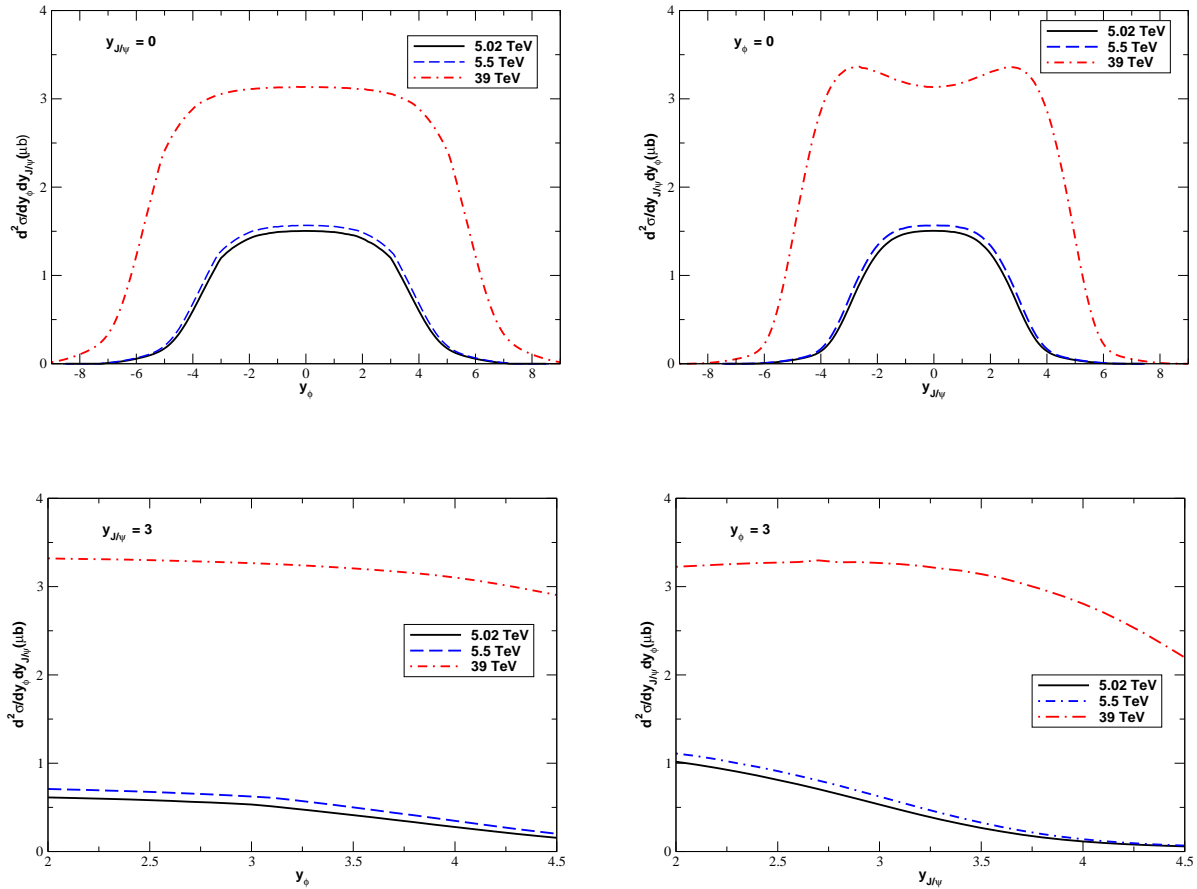

\begin{tabular}{ccc}
\includegraphics[scale=0.31]{dsigdy2.eps} & \,\,\,\,\,\,\,\,\,\,\, &
\includegraphics[scale=0.31]{dsigdy_phi0.eps} \\
\, & \, & \, \\
\, & \, & \, \\
\, & \, & \, \\
\includegraphics[scale=0.31]{rapidez_LHCb2.eps}  & \,\,\,\,\,\,\,\,\,\,\, &
\includegraphics[scale=0.31]{rapidez_LHCb_phi0.eps}  
\end{tabular}
\caption{Rapidity distributions for the $\phi J/\Psi$ production in ultraperipheral $PbPb$ collisions at the LHC and FCC energies considering the rapidity ranges covered by the ALICE (upper panels) and LHCb (lower panels) detectors. In the left (right) panels, the rapidity of the $J/\Psi$ ($\phi$) meson is assumed constant and the dependence on the rapidity of the $\phi$ ($J/\Psi$) meson is analyzed.}
\label{Fig:rapidity}
\end{figure}

In Fig. \ref{Fig:rapidity}  we present our predictions for the rapidity distributions considering the $\phi J/\Psi$  production through the double scattering mechanism in ultraperipheral $PbPb$  collisions and assuming different center - of - mass energies. We will assume a fixed rapidity for one of the vector mesons and analyze the dependence on the rapidity of the other meson. In the upper (lower) panels we present the results for the rapidity range covered by the ALICE (LHCb) detector. As demonstrated in Ref. \cite{DSM}, the rapidity distribution for a final state composed by different mesons is asymmetric, being wider for a lighter meson. Such conclusion is also verified in the results presented in Fig. \ref{Fig:rapidity} when we compare the left and right panels. As expected, the distributions  become wider and increase in normalization with the center - of - mass energy.

{

Two comments are in order before summarizing our main results and conclusions. First, as discussed in detail in Ref. \cite{run2}, the predictions for the single vector meson photoproduction in UPHIC are  dependent on the models assumed for the dipole - proton scattering amplitude ${\cal{N}}_p$ and for the overlap function $(\Psi^{V*}\Psi)_{T}$, with the uncertainty on the predictions being non - negligible. Such uncertainties have a direct impact on our predictions for the double vector meson photoproduction. In particular, one has verified that our predictions for central (forward) rapidities are modified by $\approx 30 \%$ (17 \%), if the alternative models discussed in Ref. \cite{run2} are considered as input in our calculations. Another important comment is that, in recent years, several groups have improved the treatment of exclusive processes 
in the dipole approach, by estimating higher order corrections for the evolution of the forward dipole - target scattering amplitude and  photon impact factor, as well as by improving the description of the vector meson wave function (See, e.g. Refs. \cite{Lappi:2020ufv,Beuf:2020dxl,Mantysaari:2021ryb,
Mantysaari:2022bsp,Mantysaari:2022kdm}).  Such results indicate that the NLO corrections are numerically important, but their effect can be partially captured when the initial condition for the small-$x$ evolution of the dipole amplitude is fitted to the structure function data. Moreover, the NLO predictions for the vector meson production at HERA are similar to those obtained using the phenomenological dipole models, in particular to those derived using the b-CGC model considered in this paper. Therefore, in principle, we do not expect a large modification of our predictions if the NLO corrections are taken into account. Surely, a more detailed comparison between our results and those derived at NLO is an important next step, which we plan to perform in a future study.    
}

\section{Summary}

As a summary, one has that in recent years, several studies have demonstrated that the analysis of the exclusive vector meson photoproduction in hadronic colliders is a promising way  to constrain the 
QCD dynamics at high energies. In addition, some groups have pointed out that for ultraperipheral heavy ion collisions, where the cross sections for the single production are large,  the probability of having multiple interactions in a same event is non - negligible, and can also be used to improve our understanding of the process. In particular, in Ref. \cite{DSM}, we have presented a detailed analysis of the $J/\Psi J/\Psi$, $\rho \rho$ and $\rho J/\Psi$ production in $PbPb/pPb/pp$ collisions through the double scattering mechanism. In that reference, we have demonstrated that  the DSM contribution  is dominant in $PbPb$ collisions in comparison to the double vector meson production through $\gamma \gamma$ interactions. In this paper, we have extended the analysis for the associated production of  $\phi$ and $J/\Psi$ mesons in $PbPb$ collisions for the LHC and FCC energies and presented, for the first time, the predictions for the total cross sections and rapidity distributions considering the phase space covered by the ALICE and LHCb Collaborations. As a by-product, we also have presented the results for the double $\phi$ production. Our study has been strongly motivated by the analysis of the $\phi J/\Psi$ final state that is being performed by the LHCb Collaboration, which should be released in the forthcoming years. We predict large values for the total cross sections and for the number of events in future runs of the LHC and FCC, which indicate that the study of this final state is, in principle, feasible. Considering the results presented in this paper and those in Ref. \cite{DSM}, we strongly motivate a future investigation of the double vector meson production in ultraperipheral heavy ion collisions in order to probe the double scattering mechanism and improve our understanding of the QCD dynamics at high energies.

\begin{acknowledgments}
V.P.G. thanks Murilo S. Rangel (UFRJ/Brazil and LHCb Collaboration) for useful discussions. This work was partially supported by CNPq, CAPES, FAPERGS and  INCT-FNA (Process No. 464898/2014-5).  V.P.G. was also partially supported by the CAS President's International Fellowship Initiative (Grant No.  2021VMA0019).

\end{acknowledgments}

\hspace{1.0cm}

\end{document}